\documentclass{cupconf}
\usepackage{graphicx}%%line added by slt%%

  \checkfont{eurm10}
  \iffontfound
    \IfFileExists{upmath.sty}
      {\typeout{^^JFound AMS Euler Roman fonts on the system,
                   using the 'upmath' package.^^J}%
       \usepackage{upmath}}
      {\typeout{^^JFound AMS Euler Roman fonts on the system, but you
                   dont seem to have the}%
       \typeout{'upmath' package installed. cupconf.cls can take advantage
                 of these fonts,^^Jif you use 'upmath' package.^^J}%
      }
  \else
  \fi

  \checkfont{msam10}
  \iffontfound
    \IfFileExists{amssymb.sty}
      {\typeout{^^JFound AMS Symbol fonts on the system, using the
                'amssymb' package.^^J}%
       \usepackage{amssymb}%
         \let\leq=\leqslant
         
      }{}
  \fi

  \IfFileExists{amsbsy.sty}
    {\typeout{^^JFound the 'amsbsy' package on the system, using it.^^J}%
     \usepackage{amsbsy}}
    {}

\newsavebox{\astrutbox}
\sbox{\astrutbox}{\rule[-5pt]{0pt}{20pt}}

\title[Estimating Black Hole Spin]{Estimating the Spins of\break
  Stellar-Mass Black Holes}

\author[J.\ McClintock, R.\ Narayan \& R. Shafee] {J\ls E\ls F\ls F\ls R\ls
  E\ls Y\ns E.\ns M\ls C\ls C\ls L\ls I\ls N\ls T\ls O\ls C\ls K\break
  R\ls A\ls M\ls E\ls S\ls H\ns N\ls A\ls R\ls A\ls Y\ls A\ls N\break
  \and R\ls E\ls B\ls E\ls C\ls C\ls A\ns S\ls H\ls A\ls F\ls E\ls E}%

\affiliation{Harvard-Smithsonian Center for Astrophysics, 60 Garden St., 
Cambridge, MA 02138, USA}

\pubyear{2007}
\volume{538}
\pagerange{1--10}
\date{?? and in revised form ??}
\begin{document}

\def\msun{$M_{\odot}$}
\def\mdot{$\dot M$}
\def\zz{$\a_*$}
\def\risco{$R_{\rm ISCO}$}

\maketitle

\begin{abstract}
We describe a program that we have embarked on to estimate the spins
of stellar-mass black holes in X-ray binaries.  We fit the continuum
X-ray spectrum of the radiation from the accretion disk using the
standard thin disk model, and extract the dimensionless spin parameter
$a_*=a/M$ of the black hole as a parameter of the fit.  We have
obtained results on three systems, 4U 1543-47 ($a_*=0.7-0.85$), GRO
J1655-40 ($0.65-0.8$), and GRS 1915+105 ($0.98-1$), and have nearly
completed analysis of two additional systems.  We anticipate expanding
the sample of spin estimates to about a dozen over the next several
years.
\end{abstract}

\firstsection % if your document starts with a section,
              % remove some space above using this command.

\section{Introduction}

The first black hole (BH), Cygnus X-1, was identified and its mass
estimated in 1972.  We now know of about 40 stellar-mass black holes
in X-ray binaries in the Milky Way and neighboring galaxies.  The
masses of 21 of these, which range from $\sim 5-15$\msun, have been
measured by observing the dynamics of their binary companion stars
(Remillard \& McClintock 2006; Orosz et al.\ 2007).  In addition, it
has become clear that virtually every galaxy has a supermassive black
hole with $M \sim 10^6 - 10^{10}$\msun~in its nucleus.  A few dozen of
these supermassive BHs have reliable mass estimates, which have been
obtained via dynamical observations of stars and gas in their vicinity
(Begelman 2003).

With many mass measurements now in hand, the next logical step is to
measure spin.  This would mark a major milestone since, once we have
both a BH's mass and spin, we will have achieved a complete
description of the object.  Furthermore, spin is arguably the more
important parameter.  Mass simply supplies a scale, whereas spin
changes the geometry and fundamentally conditions the ways in which a
BH interacts with its environment.  

Unfortunately, spin is much harder to measure than mass.  The effects
of spin are revealed only in the regime of strong gravity close to the
hole, where the sole probe available to us is the accreting gas.
Thus, we must make accurate observations of the radiation emitted by
the inner regions of the accretion disk, and we must have a reliable
model of the emission.  Until recently, there was no credible
measurement of BH spin.

The situation has changed within the last couple of years.  Following
up on the pioneering work of Zhang, Cui \& Chen (1997), the first
breakthrough came with estimates of the spin parameter $a_*\equiv a/M$
reported by our group (see Table 1) for three stellar-mass BHs (Shafee
et al.\ 2006; McClintock et al.\ 2006): GRO J1655--40, 4U 1543--47,
and GRS 1915+105.  These spin estimates were obtained by modeling the
continuum X-ray spectrum from the accretion disk surrounding the BHs.
Following our work, the spin of a supermassive BH was estimated by an
independent method, modeling the profile of the Fe K line (Brenneman
\& Reynolds 2006).

This paper is organized as follows.  In \S2 we describe the
continuum-fitting method and comment on our efforts to establish our
methodology.  In \S3 we review the extensive evidence for the
existence of a stable inner accretion-disk radius, which provides a
strong empirical foundation for the continuum-fitting method of
determining spin.  The importance of measuring spin is briefly
described in \S4.  In \S5 we discuss work in progress and future
prospects, and we offer our conclusions.

\begin{center}
\begin{table}

\begin{tabular}{lccl}

\multicolumn{4}{c}{TABLE 1} \\  \\ \multicolumn{4}{c}{Spin
Estimates of Stellar-Mass Black Holes} \\

\hline \hline
\multicolumn{1}{l}{BH Binary System}
&\multicolumn{1}{c}{$M/M_\odot$}&\multicolumn{1}{c}{$a_*$}
&\multicolumn{1}{l}{Reference} \\
\hline 4U 1543--47 &$~~~~9.4\pm1.0~~~~$ &$~~~~0.7 - 0.85~~~~$ 
&Shafee et al. (2006) \\
\hline GRO J1655--40 &$~~~~6.30\pm0.27~~~~$ &$~~~~0.65 - 0.8~~~~$ 
&Shafee et al. (2006) \\
\hline GRS 1915+105 &$~~~~14\pm4.4~~~~$ &$~~~~0.98 - 1~~~~$ 
&McClintock et al. (2006) \\
\hline\hline
\end{tabular}

\end{table}
\end{center}

\section{The Method: Fitting the X-ray Continuum Spectrum}

A definite prediction of relativity theory is the existence of an
innermost stable circular orbit (ISCO) for a test particle orbiting a
BH.  Once a particle is inside this radius, it suddenly plunges into
the hole.  Gas in a geometrically thin accretion disk has negligible
pressure support in the radial direction and behaves for many purposes
like a test particle.  Thus, the gas spirals in (through the action of
viscosity) via a series of nearly circular orbits until it reaches the
ISCO, at which point it plunges into the BH.  In other words, the disk
is effectively truncated at an inner edge located at the ISCO.

In our method, we estimate the radius of the inner edge of the disk by
fitting the X-ray continuum spectrum and identify this radius with
$R_{\rm ISCO}$, the radius of the ISCO.  Since the dimensionless ratio
$\xi \equiv R_{\rm ISCO}/(GM/c^2)$ is solely a monotonic function of
the BH spin parameter $a_*$ (Fig.\ 1), knowing its value allows one
immediately to infer the BH spin parameter $a_*$.  The variations in
\risco~are large: e.g., for a BH of 10\msun, \risco~ranges from 90 km
down to 15 km as $a_*$ increases from 0 to unity, which implies that
we should in principle be able to estimate $a_*$ with good precision.
%We now discuss the theoretical basis of
%our model and its requirements and application.

\begin{figure}
\begin{center}
\includegraphics[width=3.5in]{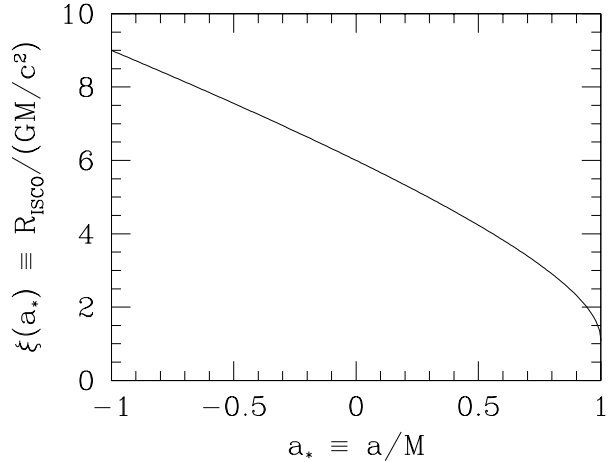}
\end{center}
\vspace{-1.0cm}
\caption{Shows the dependence of the quantity, $\xi =R_{\rm
ISCO}/(GM/c^2)$, on the BH spin parameter, $a_* \equiv a/M = cJ/GM^2$,
where $M$ and $J$ are the mass and angular momentum of the BH (Shapiro
\& Teukolsky 1983).  The spin parameter is restricted to the range $-1
\leq a_* \leq 1$; negative values correspond to the BH
counter-rotating with respect to the orbit.}
\end{figure}

The idealized thin disk model (Novikov \& Thorne 1973) describes an
axisymmetric radiatively-efficient accretion flow in which, for a
given BH mass $M$, mass accretion rate $\dot M$ and BH spin parameter
$a_*$, we can calculate precisely the total luminosity of the disk,
$L_{\rm disk} = \eta\dot Mc^2$, where the radiative efficiency factor
$\eta$ is a function only of $a_*$, as well as the profile of the
radiative flux $F_{\rm disk}(R)$ emitted as a function of radius $R$.
Moreover, the accreting gas is optically thick, and the emission is
thermal and blackbody-like, making it straightforward to compute the
spectrum of the emission.  Most importantly, as discussed above, the
inner edge of the disk is located at the ISCO of the BH space-time.
By analyzing the spectrum of the disk radiation and combining it with
knowledge of the distance $D$ to the source and the mass $M$ of the
BH, we can obtain $a_*$.  This is the principle behind our method of
estimating BH spin, which was first described by Zhang et al.\ (1997;
see also Gierli\'nski, Maciolek-Niedz\'wiecki \& Ebisawa 2001).

In practice, as we describe below, the method involves fitting X-ray
spectral data to a fully relativistic model of the disk emission and
obtaining $a_*$ as a fit parameter.  However, one can understand the
method qualitatively by noting that it effectively seeks to measure
the radius of the ISCO.  Before discussing how this is done, we remind
the reader how one measures the radius $R_*$ of a star.  Given the
distance $D$ to the star, the radiation flux $F_{\rm obs}$ received
from the star, and the temperature $T$ of the continuum radiation, the
luminosity of the star is given by
\begin{equation}
L_* = 4\pi D^2 F_{\rm obs} = 4\pi R_*^2 \sigma T^4.
\end{equation}
Thus, from $F_{\rm obs}$ and $T$, we can obtain the solid angle
$\pi(R_*/D)^2$ subtended by the star, and if the distance is known, we
immediately obtain the stellar radius $R_*$.  Of course, for accurate
results we must allow for limb darkening and other non-blackbody
effects in the stellar emission by computing a stellar atmosphere
model.

The same principle applies to an accretion disk, but with some
differences.  First, since $F_{\rm disk}(R)$ varies with radius, the
radiation temperature $T$ also varies with $R$.  But the precise
variation is known for the idealized thin disk, so it is easily
incorporated into the model.  Second, since the bulk of the emission
is from the inner regions of the disk, the effective area of the
radiating surface is directly proportional to the square of the disk
inner radius, $A_{\rm eff} = C R_{\rm ISCO}^2$, where the constant $C$
is known.  Third, the observed flux $F_{\rm obs}$ depends not only on
the luminosity and the distance, but also on the inclination $i$ of
the disk to the line-of-sight\footnote{We assume that the spin of the
BH is approximately aligned with the orbital angular momentum vector
of the binary; there is no strong contrary evidence despite the
often-cited examples of GRO J1655-40 and SAX J1819.3-2525 (see \S2.2
in Narayan \& McClintock 2005).}.  Allowing for these differences, one
can write a relation for the disk problem similar in spirit to
eq. (2.1), but with additional geometric factors that are readily
calculated from the disk model.  Therefore, in analogy with the
stellar case, given $F_{\rm obs}$ and a characteristic $T$ (from X-ray
observations), one obtains the solid angle subtended by the ISCO: $\pi
\cos i\, (R_{\rm ISCO}/D)^2$.  If we know $i$ and $D$, we obtain
$R_{\rm ISCO}$, and if we also know $M$, we obtain $a_*$.  This is the
basic idea of the method.

We note in passing that for the method to succeed it is essential to
have accurate measurements of the BH mass $M$, inclination of the
accretion disk $i$, and distance $D$ as inputs to the
continuum-fitting process (Shafee et al.\ 2006; McClintock et al.\
2006).  This dynamical work is not discussed here, although roughly
half of our total effort is directed toward securing these dynamical
data (e.g., Orosz et al.\ 2007).

Given accurate information on $M$, $i$ and $D$, there are three main
issues that must be dealt with before applying the method: 

\noindent (1) We must carefully trace rays from the surface
of the orbiting disk to the observer in the Kerr metric of the
rotating BH in order to compute accurately the observed flux and
spectrum.  To this end, our group has developed a model called {\sc
kerrbb} (Li et al.\ 2005) which has been incorporated into XSPEC
(Arnaud 1996) and is now publicly available for fitting X-ray data.

\noindent (2) We need an accurate model of the disk atmosphere for
computing the spectral hardening factor $f$ (see \S4).  We use the
advanced models of our collaborator Shane Davis (Davis et al.\ 2005)
and this element is thus well in hand.  Specifically, we have computed
tables of $f$ versus $L/L_{\rm Edd}$ for a wide range of models.
Further, we have incorporated these into a new version of {\sc kerrbb}
dubbed {\sc kerrbb2} (McClintock et al.\ 2006), which allows us to fit
directly for the spin parameter $a_*$ and the mass accretion rate
\mdot.

\noindent (3) Most importantly, the accretion disk around
the BH must be well described by the standard geometrically-thin and
optically-thick disk model, whose validity is assumed by {\sc kerrbb}
and {\sc kerrbb2}.  To ensure this, we restrict our attention strictly
to observations in the thermal state (optically thick emission) and
limit ourselves to luminosities below 30\% of the Eddington limit
(McClintock et al.\ 2006; Shafee, Narayan \& McClintock 2007).

\noindent Beyond these three issues, we must ultimately push
theory to its limits in order to understand accretion processes near
the ISCO and to obtain the most accurate model of $F_{\rm disk}(R)$
that can be achieved (see \S3).

For a full description of the mechanics of our current
continuum-fitting methodology, we refer the reader to \S4 in
McClintock et al.\ (2006).  In brief, we first select
rigorously-defined thermal-state X-ray data (\S4; Remillard \&
McClintock 2006).  We then fit the broadband X-ray continuum spectrum
using our fully relativistic model of a thin accretion disk ({\sc
kerrbb2}) in Kerr space-time, which includes all relativistic effects
(Li et al.\ 2005) and an advanced treatment of spectral hardening
(\S4; Davis et al.\ 2005).  The model also includes self-irradiation
of the disk (``returning radiation''), the effects of limb darkening,
and the effect of a torque of any magnitude at the inner edge of the
disk, although our published results are based on zero torque, which
is justified in Shafee et al. (2007).  As noted above, our new hybrid
code {\sc kerrbb2} allows us to fit directly for the two parameters of
interest: the spin $a_*$ and the mass accretion rate $\dot M$.  Using
the known radiative efficiency factor $\eta$ of the disk for a given
$a_*$, and the fitted value of $\dot M$, we compute for each
observation the Eddington-scaled luminosity, $L/L_{\rm Edd}$, and
consider only those observations for which $L/L_{\rm Edd} < 0.3$ (\S3;
Shafee et al.\ 2007).  Finally, we present our results in the form of
plots of $a_*$ versus log($L/L_{\rm Edd})$.

As an example, Figure 2 shows our results on GRS 1915+105 (McClintock
et al. 2006).  Over the luminosity range $L/L_{\rm Edd} < 0.3$, the
data are consistent with a single value of $a_*$ close to unity.
Allowing for statistical errors and uncertainties in the input values
of $M$, $i$ and $D$, we estimate $a_*$ to lie in the range $0.98-1$
(Table 1).  For luminosities closer to Eddington, the $a_*$ estimates
obtained using our method are lower, as also found by Middleton et
al. (2006), who analyzed three observations with luminosities between
$0.4L_{\rm Edd}$ and $1.4L_{\rm Edd}$ (cf. McClintock et al. 2006,
Fig. 12).  Neither the cause for the decrease nor its magnitude are
presently understood.  However, it is not surprising that our model,
which assumes a geometrically thin accretion disk, should fail at
luminosities close to Eddington when the disk is likely to be very
thick.

\begin{figure}
\begin{center}
\includegraphics[width=3.5in]{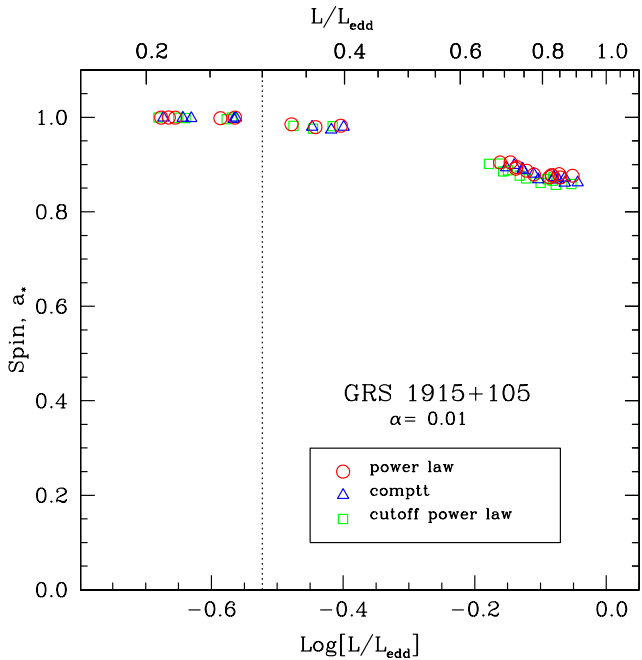}
\end{center}
\caption{Shows the estimated spin parameter $a_*$ of the BH in GRS
1915+105, as a function of the Eddington-scaled luminosity $L/L_{\rm
Edd}$.  The spectral data were analyzed using {\sc kerrbb2} combined
with three different models of the high energy Comptonized radiation
(shown by different symbols).  For $L/L_{\rm Edd} < 0.3$ (to the left
of the vertical dotted line), all the estimates of $a_*$ are
consistent with a value nearly equal to unity.  The result is
insensitive to the precise Comptonization model used in the analysis.
(Taken from McClintock et al. 2006).}
\end{figure}

The results we published on 4U1543--47 and GRO J1655--40 in Shafee et
al. (2006) were obtained with {\sc kerrbb}.  We have re-analyzed the
same data using {\sc kerrbb2}, which gives a slightly larger range of
uncertainty for the derived values of $a_*$.  The spin values listed
in Table 1 correspond to the more recent analysis.

\section{Establishing the Continuum-Fitting Method}

Given our straightforward methodology and our in-depth experience in
determining the spins of three BHs, we are confident that we can
achieve our goal of amassing a total of a dozen or so measurements of
BH spin during the next 3--4 years.  Equally important, however, are
our efforts to demonstrate that our methodology is sound.  
%Not surprisingly, our knowledge of accretion disk physics close to a BH is
%still incomplete.  
The largest systematic error in the BH spin estimates reported so far
arise from uncertainties in the validity of the disk model we employ.
Thus, it is obviously crucial to pursue detailed theoretical studies
of the physics of BH accretion flows near the ISCO.

Recently, we obtained encouraging preliminary results (Shafee et
al. 2007) based on a hydrodynamic study showing that the errors in our
spin estimates due to viscous torque and dissipation near the ISCO are
quite modest for disk luminosities $\lesssim$ 30\% of the Eddington
limit.  This is the luminosity limit that we had already adopted in
our earlier work (McClintock et al.\ 2006).  We are presently working
to extend these hydro models to full GR MHD, where magnetic stresses
may possibly cause important deviations from the standard thin disk
model (e.g., Krolik 1999; Gammie 1999; Krolik \& Hawley 2002).

In addition to this fundamental theoretical work, we are engaged in a
broader effort to assess all scenarios that can ultimately impact upon
our estimates of BH spin.  Two examples: (1) With J.\ C.\ Lee, we are
examining the possible effects of warm absorbers (i.e., photoionized
gas) on our spin estimates via an analysis of HETG grating spectra; and
(2) we are in the process of making a stringent test of our spin model
by obtaining a VLBA parallax distance and improved radial velocities for
the microquasar GRS 1915+105 (see \S6.4 in McClintock et al.\ 2006).

\section{A Basis for Optimism}

Among the several spectral states of accreting BHs, the {\it thermal
state} (see Table 2 in Remillard \& McClintock 2006), formerly known
as the high soft state, is central to the work proposed here.  A
feature of this state is that the X-ray spectrum is dominated by a
soft blackbody-like component which is emitted by (relatively) cool
optically-thick gas in the accretion disk.  In addition, there is a
minor nonthermal tail component of emission, which probably originates
from a hot optically-thin corona.  In practice, this poorly-understood
Comptonized component of emission contributes $\lesssim~10$\% of the
flux in a 2--20 keV band (e.g., {\it RXTE}) and an even much smaller
fraction in an 0.5-10 keV band (e.g., {\it ASCA} and {\it Chandra}),
which captures nearly all of the $\sim 1$ keV thermal spectrum.  Thus,
the only spectra we consider -- thermal-state spectra -- are largely
free of the uncertain effects of Comptonization (e.g., Fig. 2).  These
observed spectra are believed to match very closely the classic thin
accretion disk models of the early 1970s (Shakura \& Sunyaev 1973;
Novikov \& Thorne 1973).

There is a long history of evidence suggesting that fitting the X-ray
continuum is a promising approach to measuring BH spin.  This history
begins in the mid-1980s with the simple non-relativistic multicolor disk
model (Mitsuda et al.\ 1984; Makishima et al.\ 1986), which returns the
color temperature $T_{\rm in}$ at the inner-disk radius $R_{\rm in}$.
In their review paper on BH binaries, Tanaka \& Lewin (1995) summarize
examples of the steady decay (by factors of 10--100) of the thermal flux
of transient sources during which $R_{\rm in}$ remains quite constant
(see their Fig.\ 3.14).  They remark that the constancy of $R_{\rm in}$
suggests that this fit parameter is related to the radius of the ISCO.
More recently, this evidence for a constant inner radius in the thermal
state has been presented for a number of sources via plots showing that
the bolometric luminosity of the thermal component is approximately
proportional to $T_{\rm in}^4$ (Kubota, Makishima, \& Ebisawa 2001;
Kubota \& Makishima 2004; Gierli\'nski \& Done 2004; Abe et al.\ 2005;
McClintock et al.\ 2007).

We now demonstrate that the case for the constancy of the inner disk
radius is further strengthened if one considers the effects of
spectral hardening, which we determine via the state-of-the-art disk
atmosphere models of Davis et al.\ (2005).  At the high disk
temperatures typically found in BH disks ($T_{\rm in} \sim 10^7$ K),
non-blackbody effects are important and one replaces $T_{\rm in}$ by
the effective temperature $T_{\rm eff} = T_{\rm in}/f$, where $f$ is a
``spectral hardening factor'' (Shimura \& Takahara 1995; Merloni,
Fabian, \& Ross 2000; Davis et al.\ 2005).  In Figure 3, we illustrate
the effects of spectral hardening on the relationship between
luminosity and temperature for two BH transients (see also Davis,
Done, \& Blaes 2006).  The figure extends results that are presented
in Figure~8 in McClintock et al.\ (2007).  The top two panels show the
Eddington-scaled luminosities of the two BH transients during their
entire outburst cycles.  The bold plotting symbols denote the
rigorously-defined thermal-state data (see Table 2 in Remillard \&
McClintock 2006).  In the lower panels, we consider only these
thermal-state spectral data, and we ignore the remaining data that are
strongly Comptonized and for which the models are very uncertain.

\begin{figure}
\includegraphics[width=\textwidth]{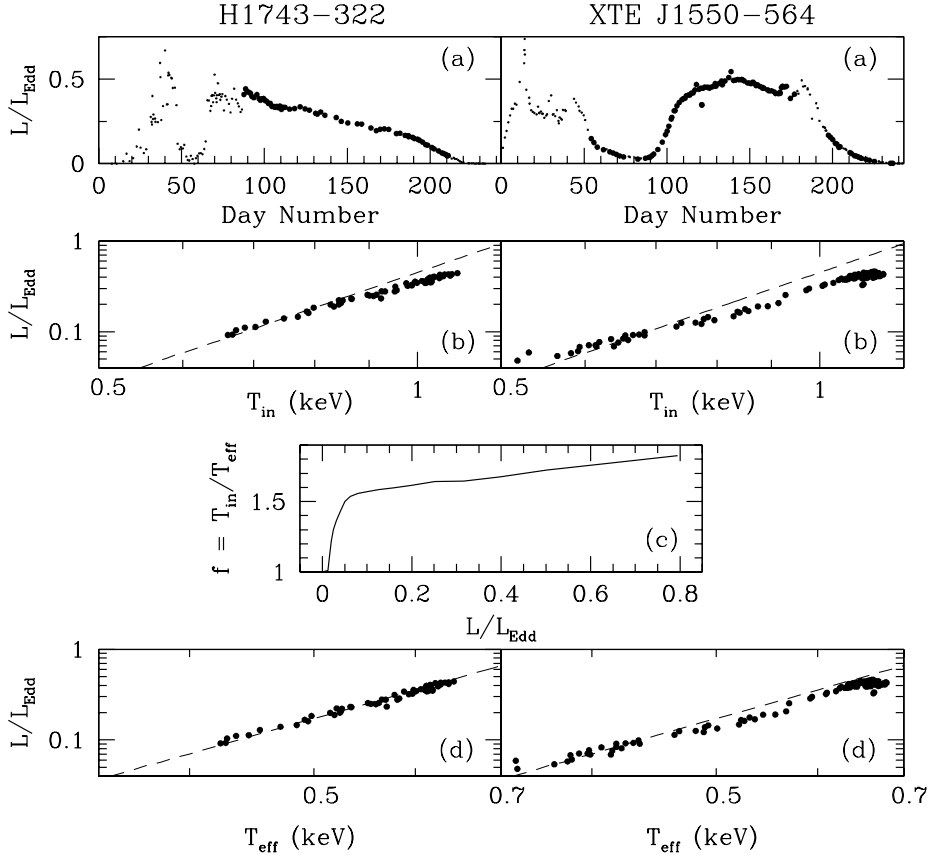}%%slt changed width to \textwidth%%
\caption{Evidence for the constancy of the inner disk radius and an
illustration of the effects of spectral hardening.  Shown are
thermal-state data collected for H1743-322 in 2003 and XTE J1550-564
in 1998--1999 in hundreds of pointed observations using the {\it RXTE}
PCA detector (McClintock et al.\ 2007). ($a$) The evolution of the
luminosities of the two transients throughout their complete 8-month
outburst cycles.  The luminosities are scaled to the Eddington limit;
for mass and distance estimates, see McClintock et al. (2007).  ($b$)
Luminosity versus the color temperature; the log-log slope of the
dashed line is 4.  ($c$) The spectral hardening factor $f \equiv
T_{\rm in}/T_{\rm eff}$ versus luminosity computed from the disk
atmosphere model of Davis et al.\ (2005) using {\sc bhspec} in XSPEC
(Arnaud 1996).  This model was computed for a PCA response matrix in
the 2--20 keV band, $M = 10$\msun~and $i = 70^{\circ}$ (McClintock et
al.\ 2007), and $a_* = 0.5$.  The model depends only weakly on the
assumed value of the spin parameter.  ($d$) Luminosity versus the
effective temperature $T_{\rm eff} = T_{\rm in}/f$, derived from the
model results shown in panel $c$.  Note how the data here hug the
dashed $T^4$ line much more closely than in panels $b$.}
\end{figure}

Panels $b$ show plots of Eddington-scaled luminosity versus the color
temperature $T_{\rm in}$; the dashed lines show an $L/L_{\rm Edd}
\propto T_{\rm in}^4$ relation (McClintock et al. 2006).  Note that
the observed luminosity rises more slowly than $T_{\rm in}^4$, which
appears to suggest that $R_{\rm in}$ is not constant.  Panel $c$ shows
an appropriate model of the spectral hardening factor $f$ as a
function of luminosity.  Using this relationship, we replotted the
luminosity data shown in panels $b$ versus $T_{\rm eff}$, thereby
obtaining the results shown in panels $d$.  Here one finds that the
luminosity is closely proportional to $T_{\rm eff}^4$, which provides
strong evidence for the presence of a {\it stable inner disk radius}.
Obviously, this non-relativistic analysis cannot provide a secure
value for the radius of the ISCO nor even establish that this stable
radius is the ISCO.  Nevertheless, the presence of a fixed radius
indicates that the continuum-fitting method is a well-founded approach
to measuring BH spin.

\section{Importance of Measuring Spin}

In order to model the ways that an accreting BH can interact with its
environment, one must know its spin.  For example, the many proposals
relating relativistic jets to BH spin (Blandford \& Znajek 1977; Meier
2003; McKinney \& Gammie 2004; Hawley \& Krolik 2006) will remain mere
speculation until sufficient data on BH spins have been amassed and
models are tested and confirmed.  Likewise, measurements of spin are
comparably important for testing stellar-collapse models of Gamma-Ray
Burst sources (Woosley 1993; MacFadyen \& Woosley 1999; Woosley \&
Heger 2006).  Knowledge of spin is also crucial for the development of
gravitational-wave astronomy, and our Shafee et al.\ (2006) paper has
already motivated the first computation of waveforms for coalescing
BHs that includes the effects of spin (Campanelli, Lousto \& Zlochower
2006).  There are several other obvious applications of spin data,
such as crucial input to models of BH formation and BH binary
evolution (Lee, Brown \& Wijers 2002; Brown et al.\ 2007) and to
models of the powerful low-frequency QPOs (1--30 Hz) and complex,
non-thermal BH states and their evolution (Remillard \& McClintock
2006).  Finally, we note that the high spins we have measured to date
were very likely imparted to these BHs during the process of their
formation (see \S6.2 in McClintock et al.\ 2006).

\section{Conclusions and Future Prospects}

We have recently completed a thorough and precise dynamical study of
the only known eclipsing BH, M33 X-7 (Pietsch et al.\ 2006), which is
the most massive stellar BH known, $M = 15.65 \pm 1.45$\msun~(Orosz et
al.\ 2007).  Furthermore, the mass of the secondary star is $M_2 =
70.0 \pm 6.9$\msun, which puts it among the most massive stars whose
masses are well-determined.  We are presently preparing a paper on the
spin of this BH based on $\sim2$ Msec of {\it Chandra} ACIS data (Liu,
McClintock, Narayan, et al.).  We are also in the process of
estimating the spin of XTE J1550--564 using {\it RXTE} PCA data, and
we anticipate estimating the spins of more than half a dozen other
stellar-mass BHs during the next 3--4 years.

An especially exciting prospect is the possibility of obtaining
independent estimates of spin via either the Fe K line profile
(Reynolds \& Nowak 2003; Brenneman \& Reynolds 2006; Miller et al.\
2007) or high-frequency ($50-450$ Hz) QPOs (T\"or\"ok et al.\ 2005;
Remillard \& McClintock 2006), which are observed for some of these
sources.  Because spin is such a critical parameter, we and many
others are planning to pursue vigorously these additional avenues, as
this will provide arguably the best possible check on our results.
Future X-ray polarimetry missions may provide yet an additional
channel for measuring spin (e.g., Connors, Stark \& Piran 1980).

We conclude with a list of questions that motivate us.  What range of
spins will we find?  Will GRS 1915+105 stand alone, or will we find
other examples of extreme spin?  As we continue to refine our models
and our measurements of $M$, $i$ and $D$, will we consistently find
values of $a_* < 1$, or will we be challenged by apparent and
unphysical values of the spin parameter that exceed unity?  Will there
be large differences in spin between the class of young, persistent
systems with their massive secondaries (M33 X-7, LMC X-1 and LMC X-3)
and the ancient transient systems with their low-mass secondaries?
What constraints will these spin results place on BH formation,
evolutionary models of BH binaries, models of relativistic jets and
gamma-ray bursts, etc.?  What will be the implications of these spin
measurements for the emerging field of gravitational-wave astronomy in
the Advanced LIGO era?  How will this new knowledge help shape the
observing programs of {\it GLAST}, {\it Black Hole Finder Probe}, {\it
Constellation-X}, and {\it XEUS}?
%%%%%%%%%%%%%%%%%%%%%%%%%%%%%%%%%%%%%%%%%%%%%%%%%%%%%%%%%%%%%%
%%%%%%%%%%%%%%%%%%%%%%%%%%%%%%%%%%%%%%%%%%%%%%%%%%%%%%%%%%%%%%


\begin{thebibliography}{}

%%%%%%%%%%%%%%%%%%%%%%%%%%%%%%%%%%%%%%%%%%%%%%%%%%%%%%%%%%%%%%
%\bibitem[]{}\textsc{} 1991
%{.}
%\textit{.} \textbf{}, xx-xx.
%%%%%%%%%%%%%%%%%%%%%%%%%%%%%%%%%%%%%%%%%%%%%%%%%%%%%%%%%%%%%%

\bibitem[]{abe05}\textsc{Abe, Y., Fukazawa, Y., Kubota, A., Kasama, \&
D., Makishima, K.} 2005
{Three Spectral States of the Disk X-Ray Emission of the Black-Hole
Candidate 4U 1630-47.}
\textit{PASJ.} \textbf{57}, 629--641.

\bibitem[]{arn96}\textsc{Arnaud, K. A.} 1996
{XSPEC: The First Ten Years.}
IN \textit{ASP Conf.\ Ser.\ 101, Astronomical Data Analysis Software and
Systems V.} (ed. G. H. Jacoby \& J. Barnes). pp.\ 17--20. ASP.

\bibitem[]{beg03} \textsc{Begelman, M. C.}, 2003
{Evidence for Black Holes.} 2003
\textit{Science.} \textbf{300}, 1898--1904.

\bibitem[]{bla77}\textsc{Blandford, R. D., \& Znajek, R. L..} 1977
{Electromagnetic Extraction of Energy from Kerr Black Holes.}
\textit{MNRAS.} \textbf{179}, 433--456.

\bibitem[]{bre06}\textsc{Brenneman, L. W., \& Reynolds, C. S.} 2006
{Constraining Black Hole Spin via X-Ray Spectroscopy.}
\textit{ApJ.} \textbf{652}, 1028--1043.

\bibitem[]{bro07}\textsc{Brown, G. E., Lee, C.-H., Moreno-Mendez, E., \&
Walter, F. M.} 2007
{Kerr Parameters a* for GRO J1655-40 and 4U 1543-47, and their
Consequences; Modeling GRS 1915+105.}
\textit{astro-ph/0612461.}

\bibitem[]{cam06}\textsc{Campanelli, M., Lousto, C. O., \& Zlochower,
Y.}  2006
{Spinning-Black-Hole Binaries: The Orbital Hang-up.}
\textit{Phys.\ Rev.\ D.} \textbf{74}, 041501(1--5).

\bibitem[]{}\textsc{Connors, P. A., Stark, R. F., \& Piran, T.} 1980
{Polarization Features of X-ray Radiation Emitted near Black Holes.}
\textit{ApJ.} \textbf{235}, 224--244.

\bibitem[]{dav05}\textsc{Davis, S. W., Blaes, O. M., Hubeny, I., \&
Turner, N. J.} 2005 {Relativistic Accretion Disk Models of High-State
Black Hole X-Ray Binary Spectra.}  \textit{ApJ.} \textbf{621}, 372--387.

\bibitem[]{dav06}\textsc{Davis, S. W., Done, C., \& Blaes, O. M.} 2006
{Testing Accretion Disk Theory in Black Hole X-Ray Binaries.}
\textit{ApJ.} \textbf{647}, 525--538.

\bibitem[]{gam99}\textsc{Gammie, C. F.} 1999 {Efficiency of Magnetized
Thin Accretion Disks in the Kerr Metric.}  \textit{ApJ.} \textbf{522},
L57--L60.

\bibitem[]{gie04}\textsc{Gierli\'nski, M., \& Done, C.} 2004
{Black Hole Accretion Discs: Reality Confronts Theory.}
\textit{MNRAS.} \textbf{347}, 885--894.

\bibitem[]{gie01}\textsc{Gierli\'nski, M., Maciolek-Niedz\'wiecki, A.,
\& Ebisawa, K.} 2001
{Application of a Relativistic Accretion Disc Model to X-ray Spectra of
LMC X-1 and GRO J1655-40.}  
\textit{MNRAS.}  \textbf{325}, 1253--1265.

\bibitem[]{haw06}\textsc{Hawley, J. F., \& Krolik, J. H.} 2006
{Magnetically Driven Jets in the Kerr Metric.}
\textit{ApJ.} \textbf{641}, 103--116.

\bibitem[]{kro99}\textsc{Krolik, J. H.} 1999 {Magnetized Accretion
inside the Marginally Stable Orbit around a Black Hole.}
\textit{ApJ.} \textbf{515}, L73--L76.

\bibitem[]{kro02}\textsc{Krolik, J. H., \& Hawley, J. F.} 2002
{Where Is the Inner Edge of an Accretion Disk around a Black Hole?}
\textit{ApJ.} \textbf{573}, 754--763.

\bibitem[]{kub04}\textsc{Kubota, A., \& Makishima, K.} 2004
{The Three Spectral Regimes Found in the Stellar Black Hole XTE
J1550-564 in Its High/Soft State.}
\textit{ApJ.} \textbf{601}, 428--438.

\bibitem[]{kub01}\textsc{Kubota, A., Makishima, K., \& Ebisawa, K.} 2001
{Observational Evidence for Strong Disk Comptonization in GRO J1655-40.}
\textit{ApJ.} \textbf{560}, L147--L150.

\bibitem[]{lee02}\textsc{Lee, C.-H., Brown, G. E., \& Wijers,
R. A. M. J.} 2002
{Discovery of a Black Hole Mass-Period Correlation in Soft X-Ray
Transients and Its Implication for Gamma-Ray Burst and Hypernova
Mechanisms.}
\textit{ApJ.} \textbf{575}, 996--1006.

\bibitem[]{lil05}\textsc{Li, L.-X., Zimmerman, E. R., Narayan, R., \&
McClintock, J. E.} 2005
{Multitemperature Blackbody Spectrum of a Thin Accretion Disk
around a Kerr Black Hole: Model Computations and Comparison with
Observations.}
\textit{ApJS.} \textbf{157}, 335--370.

\bibitem[]{mac99}\textsc{MacFadyen, A. I., \& Woosley, S. E.} 1999
{Collapsars: Gamma-Ray Bursts and Explosions in ``Failed Supernovae.''}
\textit{ApJ.} \textbf{524}, 262--289.

\bibitem[]{mak86}\textsc{Makishima, K., Maejima, Y., Mitsuda, K., Bradt,
H. V., Remillard, R. A., Tuohy, I. R., Hoshi, R., \& Nakagawa, M.} 1986
{Simultaneous X-ray and Optical Observations of GX 339-4 in an X-ray
High State.}
\textit{ApJ.} \textbf{308}, 635--643.

\bibitem[]{mcc06}\textsc{McClintock, J., E., Shafee, R., Narayan, R.,
Remillard, R. A., Davis, S. W., \& Li, L.-X.} 2006
{The Spin of the Near-Extreme Kerr Black Hole GRS 1915+105.}
\textit{ApJ.} \textbf{652}, 518--539.

\bibitem[]{mcc07}\textsc{McClintock, J. E., Remillard, R. A., Rupen,
M. P., Torres, M. A. P., Steeghs, D., Levine, A. M., \& Orosz, J. A.}
2007
{Outburst of the X-ray Nova H1743-322: Comparisons with the Black Hole
Binary XTE J1550-564.}
\textit{ApJ.} submitted. arXiv:0705.1034.

\bibitem[]{mck04}\textsc{McKinney, J. C., \& Gammie, C. F.} 2004
{A Measurement of the Electromagnetic Luminosity of a Kerr Black Hole.}
\textit{ApJ.} \textbf{611}, 977--995.

\bibitem[]{mei03}\textsc{Meier, D. L.} 2003
{The Theory and Simulation of Relativistic Jet Formation: Towards a
Unified Model for Micro- and Macroquasars.}
\textit{New Astron.\ Rev.\.} \textbf{47}, 667--672.

\bibitem[]{mer00}\textsc{Merloni, A., Fabian, A. C., \& Ross} 2000
{On the Interpretation of the Multicolour Disc Model for Black Hole
Candidates.}
\textit{MNRAS.} \textbf{313}, 193--197.

\bibitem[]{mid06}\textsc{Middleton, M., Done, C., Gierli\'nski, M., \&
Davis, S. W.} 2006 {Black hole spin in GRS 1915+105.} \textit{ApJ.}
\textbf{373}, 1004--1012.

\bibitem[]{mil07}\textsc{Miller, J. M.} 2007
{Relativistic X-ray Lines from the Inner Accretion Disks Around Black
Holes.}
\textit{ARAA.} to appear in vol.\ \textbf{45}. arXiv:0705.0540.

\bibitem[]{mit84}\textsc{Mitsuda, K., Inoue, H., Koyama, K., et al.}
1984
{Energy Spectra of Low-Mass Binary X-ray Sources Observed from TENMA.}
\textit{PASJ.} \textbf{36}, 741--759.

\bibitem[]{nar05}\textsc{Narayan, R., \& McClintock, J. E.} 2005
{Inclination Effects and Beaming in Black Hole X-Ray Binaries.}
\textit{ApJ.} \textbf{623}, 1017--1025.

\bibitem[]{nov73}\textsc{Novikov, I. D. \& Thorne, K. S.} 1973
{Black Hole Astrophysics.} 
In \textit{Blackholes.} (ed. C. DeWitt \& B. DeWitt). 
pp.\ 343--450. Gordon \& Breach.

\bibitem[]{or007}\textsc{Orosz, J. A., McClintock, J. E., Narayan, R.,
et al.}  2007 
{A Massive Stellar Black Hole Binary in the Nearby Spiral
Galaxy Messier 33.} \textit{Nature.} submitted.

\bibitem[]{pie06}\textsc{Pietsch, W., Haberl, F., Sasaki, M., Gaetz,
T. J., Plucinsky, P. P., Ghavamian, P., Long, K. S., \& Pannuti, T. G.}
2006
{ChASeM33 Reveals the First Eclipsing Black Hole X-Ray Binary.}
\textit{ApJ.} \textbf{646}, 420--428.

\bibitem[]{rem06}\textsc{Remillard, R. A., \& McClintock, J. E.} 2006 
{X-ray Properties of Black-Hole Binaries.}
\textit{ARAA.} \textbf{44}, 49--92.

\bibitem[]{rey03}\textsc{Reynolds, C. S., \& Nowak, M. A.} 2003
{Fluorescent Iron Lines as a Probe of Astrophysical Black Hole Systems.}
\textit{Phys.\ Rep.} \textbf{377}, 389--466.

\bibitem[]{sha06}\textsc{Shafee, R., McClintock, J. E., Narayan, R.,
Davis, S. W., Li, L.-X., \& Remillard, R. A.} 2006
{Estimating the Spin of Stellar-Mass Black Holes by Spectral Fitting of
the X-ray Continuum.}
\textit{ApJ.} \textbf{636}, L113--L116.

\bibitem[]{sha07}\textsc{Shafee, R., Narayan, R., \& McClintock, J. E.}
2007
{Viscous Torque and Dissipation in the Inner Regions of a Thin Accretion
Disk: Implications for Measuring Black Hole Spin.}
\textit{ApJ.} submitted. arXiv:0705.2241.

\bibitem[]{sha73}\textsc{Shakura, N.I. \& Sunyaev, R.A.} 1973
{Black Holes in Binary Systems. Observational Appearance.}
\textit{A\&A.} \textbf{24}, 337--355.

\bibitem[]{teu83}\textsc{Shapiro, S. L. \& Teukolsky, S. A.} 1983 
{Black Holes, White Dwarfs, and Neutron Stars.} Wiley.

\bibitem[]{shi95}\textsc{Shimura, T., \& Takahara, F.} 1995
{On the Spectral Hardening Factor of the X-ray Emission from Accretion
Disks in Black Hole Candidates.}
\textit{ApJ.} \textbf{445}, 780--788.

\bibitem[]{tan95}\textsc{Tanaka, Y, \& Lewin, W. H. G.} 1995
{Black Hole Binaries.}
IN \textit{X-ray Binaries.} (ed. W. H. G. Lewin, J. van Paradijs, \&
E. P. J. van den Heuvel). pp.\ 126--174. Cambridge Univ.\ Press.

\bibitem[]{tor05}\textsc{T\"or\"ok, G., Abramowicz, M. A., Kluz\'niak,
W., Stuchl\'ik, Z.} 2005
{The Orbital Resonance Model for Twin Peak kHz Quasi
periodic Oscillations in Microquasars.}
\textit{A\&A.} \textbf{436}, 1--8.

\bibitem[]{woo93}\textsc{Woosley, S. E.} 1993
{Gamma-ray Bursts from Stellar Mass Accretion Disks around Black Holes.}
\textit{ApJ.} \textbf{405}, 273--277.

\bibitem[]{woo06}\textsc{Woosley, S. E., \& Heger, A.} 2006
{The Progenitor Stars of Gamma-Ray Bursts.}
\textit{ApJ.} \textbf{637}, 914--921.

\bibitem[]{zha97}\textsc{Zhang, S. N., Cui, W., \& Chen, W.} 1997
{X-Ray Binaries: Observational Consequences.}
\textit{ApJ.} \textbf{482}, L155--L158.

\end{thebibliography}
\end{document}